\documentclass[aps]{revtex4}
\usepackage{graphicx}
\begin{document}

\title{charmed baryon in a Diquark  model}

\author{Q. W. Wang}

\email{9685233@163.com}

 \affiliation{Institute of Modern Physics, \\
Chinese Academy of Science, P.O. Box 31,\\
 Lanzhou 730000,
P.R.China}

\begin{abstract}
A diquark model is used to investigate single charmed baryons. In
this model, baryon is composed of two diquarks and an antiquark.
Masses of lowest lying states with $J^P= 1/2 ^ \pm$ are obtained.
Some masses are smaller than other theoretic predictions which
indicates that these baryons of pentaquark structure could be
relatively stable. The results also show that $\Lambda(2882)$ may
be a pentaquark with $J^P=1/2^-$ and $\Xi(3125)$ may be a
pentaquark with $J^P=1/2^+$.
\end{abstract}

\maketitle



\section{Introduction}
 Baryons containing heavy quarks have always been interesting.
 Recently many new excited charmed baryon states have been
 discovered by CLEO, BaBar, Belle and Fermilab. Heavy baryons have narrow
 widths and they are hard to produce. As products in the
 decays of heavy mesons or in hadron colliders, the cross sections to
 produce them are amall. There are no resonant production
 mechanisms as for heavy mesons. So, heavy baryons always have
 been obtained by continuum production\cite{Roberts}.
 Furthermore, non of the quantum numbers, listed in the PDG book,
are really measured, but are assignments based on quark
model\cite{yao}.

Despite these problems, the meaning to study these baryons is
important. Heavy baryons provide a laboratory to study the
dynamics of the light quarks in the environment of heavy quark,
such as their chiral symmetry\cite{cheng}. It really is an ideal
place for studying the dynamics of diquark. In these baryons, a
heavy quark can be used as a 'flavor tag' to help us to go further
in understanding the nonperturbative QCD than do the light
baryons\cite{Roberts}. Theoretically, the study of heavy baryons
has a long story\cite{Copley,Capstick,Glozman}. But, up to now,
simple and reliable estimates for the experimental quantities
regarding to the baryon spectroscopy, the production and decay
rates are still lack. So lots of work have to be done for
theorists.

 At present, only for single-charmed baryons masses of ground
states as well as many of their excitations are known
experimentally with rather good precision.  If the spectrum of the
single charmed baryons is well known, it will provide us a
framework for studying baryons with one bottom quark and help for
understanding the doubly or triply charmed baryons. In this paper,
we use the Jaffe-Wilczek \cite{JW} model to predict masses of
single charmed pentaquark states with $J^P= {\textstyle{1 \over
2}}^ \pm $ .  In the following section we introduce the diqurk
concept and the Jaffe-Wilczek model; In section 3 we give the mass
formula and our results. In the end, a discussion will be given in
section 4.

\section{Diquark and  J-W Model}

The concept of diquark appeared soon after the original papers on
quarks\cite{Gellmann,Ida,Lich}. It was used to calculate the
hadron properties. It helps us to understand hadron structure and
high energy particle reactions\cite{anselmino}. In heavy quark
effective theory, two light quarks often refer to as diquark,
which is treated as particle in parallel with quark themself.
There are several phenomenal manifestations of diquark: the
$\Sigma-\Lambda$ mass difference, the isospin $\Delta I=1/2$ rule,
the structure function ratio of neutron to proton, \emph{et
al.}\cite{Selem,Wilczek}. There are two kinds of diquark: the good
and the bad diquark. The good diquark is more favorable
energetically than the bad, which is indicated by both the
one-gluon exchange and instanton calculations. In $SU(3)_f$, the
good diquark has flavor-spin symmetry $\bar {\textbf{3} _F} \bar
{\textbf{3} _S}$ while the bad $\textbf{6} _F \textbf{6} _S$. To
give a color singlet state, both kinds of diquark have the same
color symmetry $\bar {\textbf{3}_C}$. Two diquarks obey Boson
statistyic while two quarks in a diquark obey Fermi statistic.
Jaffe and Wilczek used only the good diquark in their
paper\cite{JW}.

Jaffe and Wilczek's pentaquark is composed of two good diquarks
and an antiquark. The two diquarks combine into a color
antisymmetric $\textbf{3}_C$ and flavor symmetric $\bar
{\textbf{6}_F}$ with components: $[ud]^2$, $[us]^2$, $[ds]^2$,
$[ud][us]_+$, $[ud][ds]_+$ and $[us][ds]_+$. In the following, we
use $[qq']$ to denote a good diquark, and $(qq')$ a bad. The spin
wave function is symmetric because the diquark has spin zero. To
give a totally symmetric wave function, an orbital excitation
between the two diquarks is needed which combines the spin of
antiquark to give state $J^P={1 \over 2}^+$ and ${3 \over 2}^+$.

Two diquarks can also combine into an $SU(3)_f$ symmetric
$\textbf{3}_F$, with no orbital excitation\cite{zhang}. Take into
the antiquark into account, we can obtain states with $J^P={1
\over 2}^-$. Jaffe and Wilczek haven't considered the ${3 \over
2}^+$ and ${1 \over 2}^-$ states in their model. In summary, we
list the quantum numbers of these diquark systems in Table
\ref{quantumnumber}.

The combination of two diquarks with an antiquark gives $SU(3)_f$
multiplets $\textbf{8} \oplus {\bar{ \textbf {10}}}$  for flavor
$\bar {\textbf{6}_F}$ two diquarks combination and $\textbf{8}
\oplus {\textbf{1}}$ for $\textbf{3}_F$. By replacing an $s$ quark
with a $c$ quark, we get states of charmed or hidden charmed
baryons. In this paper, we only consider baryons with one charm
quark and with no charmed antiquark, i.e. the single charmed
baryons.

\begin{table}
\begin{center}
\begin{tabular}{r c c  c c}
\hline
quantum numbers &\textbf{F}lavor &\textbf{C}olor &\textbf{S}pin & Orbital \\

 good diquark & $\bar {\textbf{3} _a}$ & $\bar {\textbf{3}_a}$ & $s=0_a $ & $l=0_s$\\
bad diquark & $\textbf{6} _s$ & $\bar {\textbf{3}_a}$ & $s=1_s $ & $l=0_s$\\

 2 diquarks $P=+1$ & $\bar {\textbf{6} _s}$ & $ \bar{\textbf{3}} \times \bar{\textbf{3}}
 \to \textbf{3} _a$ & $s=0_s$ &
 $l=1_a$ \\
  $P= -1$& $ \textbf{3}_a$ & $ \bar{\textbf{3}} \times \bar{\textbf{3}} \to \textbf{3} _a$ & $s=0_s$ &
 $l=0_s$ \\
\hline
\end{tabular}
\end{center}
 \caption{Summary of the diquark quantum numbers. The two quarks obey Fermi statistics
  wile the diquarks do Bose. The subscripts $a$ and $s$ are antisymmetric and symmetric for short. }
\label{quantumnumber}
\end{table}

 \section{Mass Formula and Spectrum }

   A schematic mass formula of the pentaquark reads:

   \begin{equation}
M = m_{D1}  + m_{D2}  + m_q  + E_L. \\
\label{eq:eqmass}
   \end{equation}
Here, $m_{D}$ is the diquark mass, $m_q$ the antiquark mass and
$E_L$  the orbital exciting energy. Firstly, we consider the good
diquarks. Taking the $SU(2)$ isospin symmetry into account, we
need to know the diquark mass of $[ll']$, $[ls]$ and $[lc]$, with
$l$ and $l'$ being the light quarks $u$ or $d$. There is no $[sc]$
diquark, since it is obtained by substituting one $s$ to one $c$
quark and the diquark is flavor antisymmetric.  We can get the
diquark mass by adding the two quarks mass and their binding
energy. Deducing from J-W's original paper, we use the parameters
$m_{[ll]}=420$ $MeV$ and $m_{[ls]}=580$ $MeV$. And their binding
energy we show in Table \ref{quarkmass}. The binding energy of
$[lc]$ can be obtained, with a coefficient $3/4$ as in J-W's
paper, from the mass difference of $\Sigma_c(2455)$ and
$\Lambda_c(2285)$ which are composed of a diquark and an
antiquark. We see in Table \ref{quarkmass} that two quarks having
a closer mass is more tightly bound which may be indicated by the
spin-spin interaction. And the mass splitting
\begin{equation}
(ud)-[ud]>(us)-[us]>(uc)-[uc]\simeq 0
\end{equation}
 is expected.

Moreover, a generalized Chew-Frautschi formula relating baryon
mass to orbital angular momentum is
\begin{equation}
E=\sqrt{{\sigma L}}+kL^{-{1\over 4}}(m_1 ^{3 \over 2}+m_2 ^{3
\over 2})
\end{equation}
with $k\approx 1.15$ GeV $^{-{1 \over 2}}$ and  $\sigma \approx
1.1$ GeV$^2$ \cite{Selem,Wilczek}. Here, $m_1$ and $m_2$ are the
diquark and quark mass respectively. In Ref.\cite{Selem,Wilczek},
N(1680) with good diquark $[ud]$ and $\Delta (1950)$ with bad
diquark $(ud)$ are assigned to have angular momentum $L=2$, which
give the diquark mass splitting
\begin{equation}
(ud)^{3/2}-[ud]^{3/2}\simeq0.28GeV^{3/2}.
\end{equation}
 Similarly, the mass difference of
$\Sigma(2030)$ and $\Sigma(1915)$ lead to
\begin{equation}
(us)^{3/2}-[us]^{3/2}\simeq0.12GeV^{3/2}.
\end{equation}
 From these diquark mass splittings we can get the bad diquark masses. In the
end, all the parameters we used are listed in Table
\ref{quarkmass}.

\begin{table}
\begin{center}
\begin{tabular}{c  c c  c}
\hline \hline
 quark mass & $m_c$ & $m_s$ & $m_l $
\\
\hline
   & 1650 & 460 & 360
 \\
\hline \hline
 diquark energy & $ll'$ & $ls$ & $lc$
 \\
 \hline
good & 420 & 580 & 1840
\\
\hline
 bad  & 673  & 680 & 1840
\\
\hline

 binding enegy for good& 300& 240&130\\
 \hline \hline
 \end{tabular}
\end{center}

\caption{Quark mass and diquark energy in unit MeV. }
\label{quarkmass}
\end{table}

\begin{figure}

\begin{center}
\fbox{
\includegraphics[height=8cm,width=10cm]{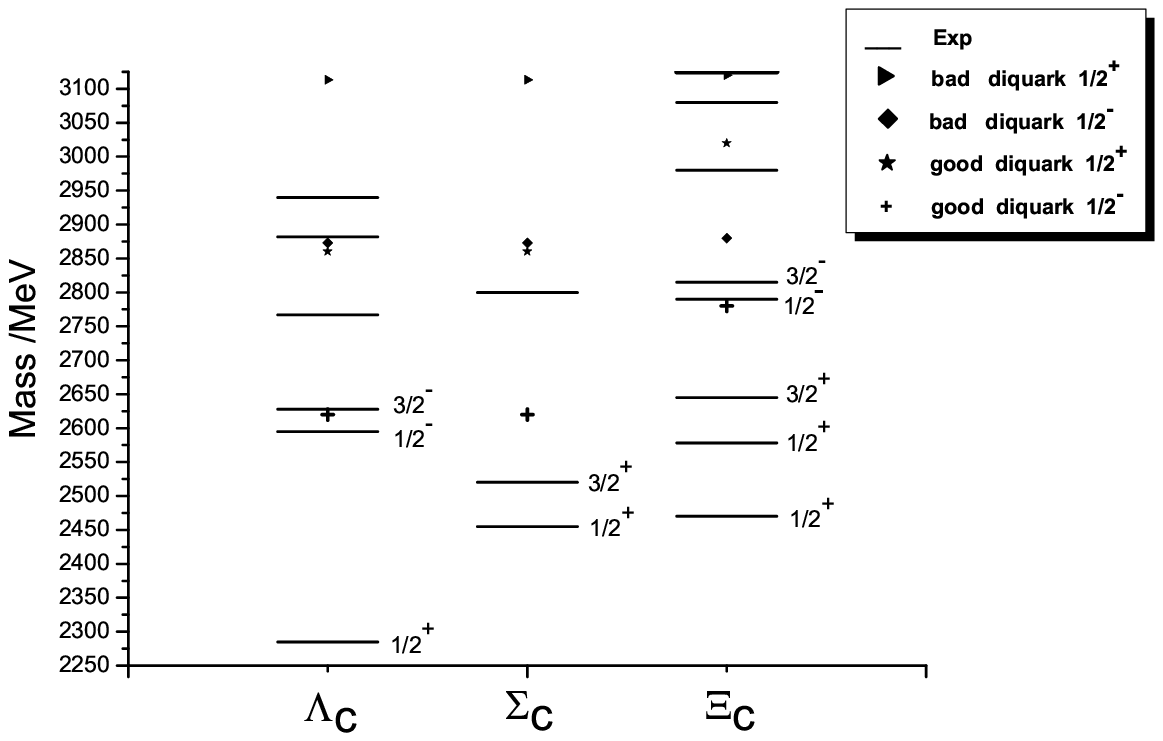}}

\caption{The single charmed baryon spectroscopy. The 'good
diquark'  labels baryons having two good diquarks, while 'bad
diquark', those having one good and one bad.}

\label{fig1}
\end{center}
\end{figure}

If we take the ideal mixing for $\textbf{8} \oplus {\bar{
\textbf{10}}}$ and $\textbf{8} \oplus {\textbf{1}}$ respectively,
the flavor assignments of charmed pentaquarks composed of  good
diquark are $[ll][lc]\bar{l}$ for $\Sigma_c $ and $\Lambda_c$ and
$[ls][lc]\bar{l}$ for $\Xi_c$. Because the mass of $[cs]$ is
unknown, we will not consider $\Omega_c$. Furthermore, in equation
(\ref{eq:eqmass}) we have not considered the splitting of the
$J^P$ $3/2^+$ and $ 1/2^+$ states. The $ 1/2^+$ is generally lower
in energy, so we take equation(\ref{eq:eqmass}) as the mass
formula for states $1/2^+$. The $J^P=1/2^-$ state with no orbital
angular momentum is a little simpler. With a substitution of
un-charmed bad diquark for good diquark, we can get more masses of
single charmed baryons. Since the bad diquark is a spin triplet,
we encounter the problem as before both for the $P=\pm1$
pentaquark. We still only take states with the lowest angular
momentum. All the masses of single charmed baryons we can get are
showed in Figure \ref{fig1}. They are $2620$, $2860$, $2873$ and
$3113$ MeV for $\Lambda_c$($\Sigma_c$), $2780$, $3020$, $2880$ and
$3120$ MeV for $\Xi_c$. For a comparison, the experimental data
are also there.

The $\Sigma_c$ and $\Lambda_c$ have the same predicted masses,
because we can't distinguish them when use formula
\ref{eq:eqmass}. The lowest predicted mass with $J^P=1/2^-$ lies
between the experimental $P=-1$ doublet. The masses of bad diquark
$(ll')$ and
 $(ls)$ are so closed which leads to baryons of the same quantum
numbers and with one bad diquark having almost the same mass. The
predicted $\Lambda_c(2873)$ to have $J^P=(1/2)^-$ is coincided
with an early conjecture\cite{Artuso} for $\Lambda_c(2882)$. For
$\Xi_c$, there is one state $\Xi_c(2790)$ near the predicted state
$\Xi_c(2780)$ with the same $J^P=1/2^-$, and $\Xi_c(3120)$ with
$J^P=1/2^+$ is very close to experimental $\Xi_c(3125)$.

The masses of $\Lambda _c(2620)$ and $\Xi_c(2780)$ of our
predictions are a little heavy than the lowest masses predicted in
Ref.\cite{Roberts,Glozman}. In their papers, different symmetry or
mix have led to more heavy results such as $\Lambda(2747,2816)$
and $\Xi(2898,2859)$, which are above our predictions of baryon
$J^P=1/2^-$ with good diquark. This means that pentaquarks may be
relatively stable than some three-quark baryons. The
$\Sigma_c(2620)$ is lower a lot than other theoretic predictions
with mass about 2700 MeV \cite{Roberts,Capstick,Glozman} which is
easy to understand because in their models   $\Sigma_c$ and
$\Lambda_c$ are well to split.

Lastly, we just simply discuss decays of these baryons.  The
combining energy of diquark with one charm quark is relatively
small. In the "fall-apart" mechanism, the dominant decay mode of
these five quark objects is decaying into a three-quark charmed
baryon and a $\pi$ meson. For example, the $J^P={\textstyle{1
\over 2}}^ -$ $\Sigma_c(\Lambda_c)$ which has a mass 2620 MeV can
decay into the $J^P={\textstyle{1 \over 2}}^ +$ $\Lambda_c(2285)$
or $ \Sigma_c(2455)$. We note that in the early paper of
Copley\cite{Copley} the decay of $\Lambda_c(2510)$
$J^P={\textstyle{1 \over 2}}^ -$ to $ \Sigma_c(2455)$ and pion is
forbidden by energy conservation.

 \section{Summary and Discussion}

In this paper, we have extended the Jaffe-Wilczek's Diquark Model
for $J^P=1/2^+$ pentaquark to dealing with charmed baryons. We
have given a spectrum of $P=\pm 1$ lowest lying charmed
pentaqurank and compared them with experimental data. We find that
some states have experimental correspondences.

The mass formula we used is just schematic. It is better to
include mass contribution from the Pauli blocking and annihilation
effects\cite{zhu}. But to quantify them is somewhat difficult.
Energy contribution from Pauli blocking is conjectured to have
relation $E_{pb}^{L=0}$ $>$ $E_{pb}^{L=1}$ which means the
two-diquark subsystem of flavor symmetry $\textbf{3}_F$ is heaver
than the one of $\bar {\textbf{6}_F}$ \cite{Wilczek,zhu}. If so,
states of $J^P=1/2^-$ will be a little heavier than our
predictions.

\bigskip
A problem?   $M_c(1650)+M_l(360)=M_D^*(2010)<M_D(2460)$ for "the
attractive one-gluon-exchange potential of a diquark in color
3-bar is a factor of 2 weaker than that of a color-singlet
qq-bar".

\section{Acknowledgments}


\begin{thebibliography}{1}
\bibitem{Roberts}W. Roberts and M. Pervin, arXiv:nucl-th/0711.2492v1
(2007).
\bibitem{yao}W. M. Yao et al. (Particle Data Group), J. Phys. G \textbf{33}  (2006)
1.
\bibitem{cheng}H. Y. Cheng, arXiv:hep-ph/0709.0958v1 (2007)
\bibitem{Copley} L. A. Copley, N. Isgur and G. Carl, Phy. Rev. D,\textbf{20}(1979)
768.
\bibitem{Capstick}S. Capstick and N. Isgur, Phy. Rev. D
\textbf{34} {1986} 2809.
\bibitem{Glozman}L. Ya. Glozman, D.O. Riska,Nucl. Phy. A \textbf{603} (1996)
326.

\bibitem{JW} R. Jaffe and F. Wilczek, Phys. Rev. Lett. \textbf{91} (2003)
232003.
\bibitem{Gellmann} M. Gell-Mann, Phys. Lett. \textbf{8} (1964)
214.

\bibitem{Ida} M. Ida, and R. Kobayashi, Prog. Theor. Phys.
\textbf{36}(1966)  846.
\bibitem{Lich} A. Lichtenberg and L.
Tassie, Phys. Rev. \textbf{155}(1967)  1601.
\bibitem{anselmino}M. Anselmino, E. Predazzi and et al., Rev. Mod.
Phy \textbf{65}(1993)  1199.
\bibitem{Selem}A. Selem and F. Wilczek, arXiv:hep-ph/0602128v1 (2006)
\bibitem{Wilczek} F. Wilczek, arXiv:hep-ph/0409168 (2004)
\bibitem{zhang}A. Zhang, Y. R. Liu and et al., arXiv:hep-ph/0403210 (2004)
\bibitem{Artuso}M. Artuso et al. [CLEO Collaboration], Phys. Rev. Lett. \textbf{86} (2001)
4479.
\bibitem{zhu}S. L. Zhu, Phy. Rev. C \textbf{70} (2004)  045201.
\end{thebibliography}
\end{document}